\documentclass[11pt,a4paper]{article}
\usepackage[english]{babel}
\usepackage{epsfig}
\usepackage{amsmath}
\usepackage{amssymb}

\font\msbmten=msbm10 scaled 1100

\def\RT0{{\rm RT}_0}

\font\msbmten=msbm10 scaled 1200
\def\Bbt#1{\msbmten#1}

\newtheorem{remark}{Remark}
\newtheorem{lemma}{Lemma}

\begin{document}

\title{ A Finite Element Framework for Option Pricing with the Bates Model
\thanks{%
Submitted for publication to \textit{Computing and Visualization in Science}}%
}
\author{E.~Miglio$^*$,C.~Sgarra$^\sharp$}
\maketitle
\date{}

\begin{abstract}
In the present paper we present a finite element approach for option pricing
in the framework of a well-known stochastic volatility model with jumps, the
Bates model. In this model the asset log-returns are assumed to follow a
jump-diffusion model where the jump component consists of a L\'{e}vy process
of compound Poisson type, while the volatility behavior is described by a
stochastic differential equation of CIR type, with a mean-reverting drift
term and a diffusion component correlated with that of the log-returns. Like
in all the L\'{e}vy models, the option pricing problem can be formulated in
terms of an integro-differential equation: for the Bates model the unknown $%
F(S, V, t)$ (the option price) of the pricing equation depends on three
independent variables and the differential operator part turns out to be of
parabolic kind, while the nonlocal integral operator is calculated with
respect to the L\'{e}vy measure of the jumps. In this paper we will present
a variational formulation of the problem suitable for a finite element
approach. The numerical results obtained for european options will be
compared with those obtained with different methods.
\end{abstract}

\begin{center}
{\small $^*$ MOX-- Modellistica e Calcolo Scientifico\\[0pt]
Dipartimento di Matematica ``F. Brioschi'' \\[0pt]
Politecnico di Milano \\[0pt]
via Bonardi 9, 20133 Milano, Italy\\[0pt]
\texttt{edie.miglio@polimi.it} }

{\small $^\sharp$ Dipartimento di Matematica ``F. Brioschi'' \\[0pt]
Politecnico di Milano \\[0pt]
via Bonardi 9, 20133 Milano, Italy\\[0pt]
\texttt{carlo.sgarra@polimi.it}}
\end{center}


\vspace*{0.5cm}


\vfill

\pagebreak


\section{Introduction}

A huge effort has been made in the last few years in order to overcome the
intrinsic limitations of the Black-Scholes model. Although it has been a
great success as a first attempt to provide an evaluation for financial
derivatives, it was soon clear that it's description of financial market
behavior was not satisfactory. Very well known observed empirical features
of the log prices were not correctly described by this model: heavy tails,
volatility clustering, aggregational gaussianity are some peculiarities that
cannot be explained on the basis of the lognormal assumption on which the
Black-Scholes model stands. The volatility smile is another relevant
phenomenon that cannot be explained on the basis of a Black-Scholes
description. Several different approaches have been exploited in order to
give a more satisfactory description of financial markets, but the main
contributions in this direction can be grouped in two different classes of
models, the so called stochastic volatility models and the models with
jumps. An extended literature is available on both kind of approaches and
they both give a more realistic description of the prices evolution in
financial markets, but separately considered they perform very well only in
some situations. While jump models can succesfully reproduce the volatility
smiles on short term maturity ranges, stochastic volatility models give a
better description of the same phenomenon on long maturity terms. This has
naturally led to the introduction of more complicated, but more realistic
models in which both features of stochastic volatility and jumps can be
present. The three more popular models in which this integration of jumps
and stochastic volatility has been performed are the BNS model introduced by
O. Barndorff-Nielsen and N. Shepard \cite{BNS1}, \cite{BNS2}, the model
introduced by Bates in \cite{Bates}, and the time-changed L\'{e}vy models
introduced by Carr, D. Madan, H. Geman and M. Yor in \cite{CGMY}. While in
the former the volatility dynamics is driven by a positive L\'{e}vy process
correlated with the jump process in the log-price of the asset, in the
latter the volatility dynamics is governed by a time-changed L\'{e}vy
process. We shall concentrate in the present work on the second model we
have just mentioned, the Bates model in which a Merton jump-diffusion model
is combined with a stochastic volatility model of the Heston type. As R.
Cont and P. Tankov have pointed out, the performance of the time changed L%
\'{e}vy models in calibrating to market option prices are usually much
better than those obtained in a BNS framework \cite{Schoutens}: in the last
model, in fact, the possible implied volatility patterns are restricted by
the requirement that the same parameter $\rho $ characterize both the jumps
in the returns and the volatility; on the other hand the calibration
performances of the Bates model are comparable to those of the time changed L%
\'{e}vy processes, "Thus the Bates model appears to be at the same time the
simplest and the most flexible of the models" \cite{CT}, pag. 495. In the
framework of option pricing via PDE (PIDE for models with jumps) several
different approaches have been exploited both for stochastic volatility
models and models with jumps. As far as finite element methods are
concerned, we just quote the papers by Y. Achdou and N. Tchou \cite{Achdou}
and by N. Hilber, A.-M. Matache and C. Schwab \cite{Schoutens} for the first
class of models and the papers by A.-M. Matache, T. von Petersdorff and C.
Schwab \cite{Matache} and by A.-M. Matache, P.-A. Nitsche and C. Schab \cite%
{Matache2} for the second. For models including both features the numerical
methods proposed until now are much less. Some authors have considered
finite-difference schemes for these models. D. Hilber, A.-M. Matache and
Schwab have provided a finite-element approach to a large class of
stochastic volatility models without jumps in \cite{Schwab}, including the
Heston model. In the present paper we shall present a finite-element
approach to option pricing in a Bates model framework. In the next section
we'll recall the basic features of the Bates model, while in section 3 we'll
provide the PIDE formulation of the option pricing in this model; in section
4 we shall present the variational formulation of the problem. In section 5
we'll describe the numerical method proposed and in section 6 we'll expose
some comments on the results obtained. In section 7 we'll outline some
conclusions and some futures perspectives of the present work.

\section{The Bates model}

In the Bates model the asset price evolution is given by:%
\begin{equation}
S_{t}=S_{0}e^{X_{t}},
\end{equation}%
where the log-returns $X$ and its volatility $Y$\ satisfy the following
stochastic differential equations:%
\begin{equation}
dX_{t}=(\alpha -{\frac{1}{2}}Y_{t})dt+\sqrt{Y_{t}}dW_{t}^{1}+dZ_{t},\quad
X_{0}=0.
\end{equation}%
\begin{equation}
dY_{t}=\xi (\eta -Y_{t})dt+\theta \sqrt{Y_{t}}dW_{t}^{2},\quad Y_{0}=y>0,
\end{equation}%
where $dW_{t}^{1}$ and $dW_{t}^{2}$ are two standard Wiener processes with
correlation $\rho$ and $dZ_{t}$ is a L\'{e}vy process of compound Poisson
type . Let's assume for the parameters the following restrictions:
\begin{equation}
\alpha \in
\mathbb{R}
,\quad -1\leq \rho \leq 1,\quad \xi >0,\quad \eta >0,\quad \theta >0
\end{equation}%
\begin{equation}
\theta ^{2}\leq 2\xi \eta .  \label{Parameters' restrict}
\end{equation}%
The last requirement is in order to assure that the volatility process $Y$
will never hit zero. We'll assume moreover $E[Z_{1}^{2}]<\infty $, this
implying that the L\'{e}vy-Khinchine representation formula for the process $%
Z$ will be of the following type:
\begin{equation}
\kappa (z)=\zeta z+\int (e^{zx}-1-zx)U(dx)
\end{equation}%
where $\zeta =E[Z_{1}]$ and $U$ is the L\'{e}vy{} measure of~$Z$. We'll
denote by $\mu (dx,dt)$ the jump measure of~$Z$ and by $\nu (dt,dx)$ its
predictable compensator. We'll have moreover $\nu (dx,dt)=U(dx)dt$. We can
write then:
\begin{equation}
Z_{t}=\zeta t+\int_{0}^{t}\int_{-\infty }^{+\infty }x(\mu -\nu )(dx,ds).
\end{equation}%
The stochastic differential equation for the price $S$ will be then:%
\begin{equation}
\begin{array}{l}
dS_{t}=(\alpha +\kappa (1))S_{t-}dt+S_{t-}\sqrt{Y_{t}}dW_{t}^{1}+ \\
\displaystyle \int_{-\infty }^{+\infty }S_{t-}(e^{x}-1)(\mu -\nu )(dx,dt).%
\end{array}%
\end{equation}

\begin{remark}
In the original model of~\cite{Bates}, the process~$Z$ is a compound Poisson
process,
\begin{equation}
Z_{t}=\sum_{i=1}^{N_{t}}J_{i},
\end{equation}%
where $N$ is a standard Poisson process with intensity~$\lambda >0$ and $%
(J_{i})_{i\geq 1}$ are iid $N(\gamma ,\delta ^{2})$, with $\gamma =\ln (1+%
\bar{k})-\delta ^{2}/2$. The corresponding cumulant function is in that case%
\begin{equation}
\kappa (z)=\lambda (e^{\gamma z+\delta ^{2}z^{2}/2}-1).
\end{equation}
\end{remark}

\begin{remark}
If $Z=0$ then we obtain Heston's stochastic volatility model from~\cite{He}.
If $\theta=0$ and $y=\eta$ we obtain Merton's jump diffusion model from~\cite%
{Merton1976} with normal jumps. Consequently we might consider the Bates
model as an extension of a Merton jump-diffusion model with stochastic
volatility, or as an extension of a Heston volatility model with jumps in
the returns.
\end{remark}

\begin{lemma}
The dynamics of the asset price process is given by
\begin{equation}  \label{dS}
\begin{array}{l}
dS_t= (\alpha+\kappa(1))S_{t-}dt+ S_{t-}\sqrt{Y_t}dW^1_t+ \\
\displaystyle \int_{-\infty}^{+\infty} S_{t-}(e^x-1)(\mu-\nu)(dx,dt).%
\end{array}%
\end{equation}
In particular if
\begin{equation}  \label{martingale}
\alpha+\kappa(1)=0
\end{equation}
the process $S$ is a local martingale.
\end{lemma}

Proof: This follows immediately from It\^{o}'s formula for general
semimartingales.\hfill $\Box $

As in other affine stochastic volatility models with and without jumps, it
is possible to obtain the characteristic function of the log-price in closed
form. This characteristic function has been calculated by D. Bates in \cite%
{Bates}; a detailed computation is provided also in \cite{CT}; it is given
by the following expression:%
\begin{eqnarray}
\Phi _{t}(u) &=&\exp \left[ t\lambda (e^{-\delta ^{2}u^{2}/2+i(\ln
(1+m)-\delta ^{2}/2)u}-1\right]  \label{cf} \\
&&\left[ \cosh \frac{\varepsilon t}{2}+\frac{\xi -i\rho \theta u}{%
\varepsilon }\sinh \frac{\varepsilon t}{2}\right] ^{-2\xi \eta /\theta ^{2}}
\\
&& \exp \left[ -\frac{(u^{2}+iu)v_{0}}{\varepsilon \coth \frac{\varepsilon t%
}{2}+\xi -i\rho \theta u}\right]
\end{eqnarray}%
where:%
\begin{equation}
\varepsilon =\sqrt{\theta ^{2}(u^{2}+iu)+(\xi -i\rho \theta u)^{2}}
\end{equation}%
Once the characteristic function of the log-price process is known in a
closed form, the valuation problem for vanilla options can be easily solved
by an FFT-related technique\ like that provided in the paper by P. Carr and
D. Madan \cite{Carr}.

A quadratic approach to option hedhing in the Bates model has been suggested
in \cite{HS}.

\section{Option pricing via PIDE approach}

By applying Ito lemma, and introducing the market price of risk $\pi $, we
obtain the following partial integro-differential equation for the price of
a European call option $C(S,y,t)$\ in the framework of the Bates model :%
\begin{equation}  \label{Pricingeq1}
\begin{array}{l}
\displaystyle \frac{\partial C}{\partial t}+(r-\kappa (1))S\frac{\partial C}{%
\partial S}+\frac{1}{2}yS^{2}\frac{\partial ^{2}C}{\partial S^{2}}+ \\[3mm]
\displaystyle [\xi (\eta -y)-\pi ]\frac{\partial C}{\partial y}+\frac{1}{2}%
\theta ^{2}y\frac{\partial ^{2}C}{\partial y^{2}}+\rho \theta yS\frac{%
\partial ^{2}C}{\partial y\partial S}+ \\[3mm]
\displaystyle \int_{-\infty }^{+\infty }\left[ C(Se^{x},y,t)-C(S,y,t)\right]
W(dx)- rC=0%
\end{array}%
\end{equation}%
with the following final condition at $t=T$:%
\begin{equation}
C(S_{T},y_{T},T)=\max \left[ S_{T}-K,0\right]
\end{equation}%
and the following boundary conditions both in $S$ and $y$:%
\begin{equation}
C(0,y,t)=0,\frac{\partial C}{\partial S}(\infty ,y,t)=1
\end{equation}%
\begin{align}
C(S,\infty ,t)&=S, \\
C(S,0,t)&=\sum_{n=0}^{\infty }e^{-\lambda t}\frac{\left( \lambda t\right)
^{n}}{n!}C_{BS}(S,t,K,\hat{\sigma}_{n},\hat{r}_{n}),
\end{align}%
where $C_{BS}(S,t,K,\hat{\sigma}_{n},\hat{r}_{n})$ are the Black-Scholes
values of the call options at time $t$ for underlying price $S$ and strike $%
K $ with parameters%
\begin{equation}
\hat{\sigma}_{n}=n\gamma ^{2}/t,
\end{equation}
\begin{equation}
\hat{r}_{n}=r+\lambda (1-e^{\gamma +\delta ^{2}/2})+n(\gamma +\delta
^{2}/2)/t
\end{equation}
By assuming lognormal jumps we have

\begin{equation}
\kappa (1)=\lambda \left( e^{\gamma -\delta ^{2}/2}-1\right),
\end{equation}
where $\gamma ,\delta ^{2}$ are the expected value and variance respectively
of the normal distribution describing the jumps' size.

The variables $S,Y$ and $t$ can assume values in the following domains: $%
s\in \lbrack 0,+\infty )$, $t\in \lbrack 0,+\infty )$ and $y\in \lbrack
0,+\infty )$ .

\begin{remark}
The market price of risk $\pi $ can be obtained in different ways in the
frame of general equilibrium models; consumption-based models give a risk
premium proportional tu $y$. In the following, we'll assume without any lost
of generality, that the market price of risk associated to the volatility is
zero. The method can be generalized to a different choice in a
straightforward way.
\end{remark}

\begin{remark}
By the commonly used substitution $S_{t}=\exp (X_{t})$ , $%
F(X_{t},y_{t},t)=C(e^{X_{t}},y_{t},t)$\ the previous equation becomes:%
\begin{equation}  \label{Pricingeq2}
\begin{array}{l}
\displaystyle\frac{\partial F}{\partial t}+(r-\kappa (1)-\frac{y}{2})\frac{%
\partial F}{\partial x}+\xi (\eta -y)\frac{\partial F}{\partial y}+ \\[3mm]
\displaystyle\frac{1}{2}y\frac{\partial ^{2}F}{\partial x^{2}}+\frac{1}{2}%
\theta ^{2}y\frac{\partial ^{2}F}{\partial y^{2}}+\rho \theta y\frac{%
\partial ^{2}F}{\partial y\partial x}+ \\[3mm]
\displaystyle \int_{-\infty }^{+\infty }\left[ F(X+u),y,t)-F(X,y,t)\right]
W(du)-rF=0%
\end{array}%
\end{equation}
\end{remark}

$W(dx)$ is the L\'{e}vy density of the jumps. In the case of gaussian jumps
for $X$ (i.e. lognormal for $S$) it will be of the following form:%
\begin{equation}
W(du)=\lambda \frac{1}{\sqrt{2\pi }\delta }\exp (\frac{(u-\gamma )^{2}}{%
2\delta ^{2}})du
\end{equation}%
It will be then a gaussian density with expected value $\gamma $ and
variance $\delta ^{2}$ multiplied by the intensity $\lambda $ of the Poisson
process.

\begin{remark}
The boundary conditions for the new unknown $F(x,y,t)$ are now the
following. For $X\in {} $:%
\begin{align}
F(-\infty ,y,t)&=0, \\
F(+\infty ,y,t)&=e^{x},  \label{Boundarycond0} \\
F(x,0,t)&=\sum_{n=0}^{\infty }e^{-\lambda t}\frac{\left( \lambda t\right)
^{n}}{n!}C_{BS}(e^{x},t,K,\hat{\sigma}_{n},\hat{r}_{n}), \\
F(x,\infty ,t)&=e^{x}.
\end{align}%
The final condition is now:%
\begin{equation}
F(X_{T},y_{T},T)=\max \left[ e^{X_{T}}-K,0\right]
\end{equation}
\end{remark}

\section{Variational Formulation}

The integro-differential equation given before can be written in the
following "divergence form":%
\begin{equation}
\begin{array}{l}
\displaystyle\frac{DF}{Dt}+\nabla\cdot\left[ \mathbf{K\nabla }F\right] + \\%
[3mm]
\displaystyle\int_{-\infty }^{+\infty }\left[ F(X+u),y,t)-F(X,y,t)\right]
W(du)-rF=0 \label{Divergence}%
\end{array}%
\end{equation}%
where the symbol $\displaystyle\frac{D}{Dt}=\frac{\partial }{\partial t}+(%
\mathbf{a}\cdot\nabla)$ denotes the total derivative, the vector $\mathbf{a}$
is given by:%
\begin{equation}
\mathbf{a:=}\left[
\begin{array}{c}
r-\kappa (1)-y/2-\rho \theta /2 \\
\xi (\eta -y)-\theta ^{2}/2%
\end{array}%
\right]
\end{equation}%
and the matrix $\mathbf{K}$ by:%
\begin{equation}
\mathbf{K:=}\left[
\begin{array}{cc}
y/2 & \rho \theta y/2 \\
\rho \theta y/2 & \theta ^{2}y/2%
\end{array}%
\right]
\end{equation}%
A variational formulation for the PDE arising in the Heston model has been
given in \cite{Schwab}, while the variational formulation for the PIDE in an
Exponential L\'{e}vy framework have been provided in\ \cite{Matache}, \cite%
{Matache2}, \cite{Zhang}, where existence and uniqueness of the solution of
the variational problem associated with the differential and
integro-differential equations respectively have been proved in suitable
weighted Sobolev spaces and detailed analyses of both localization and
discretization errors have been provided. Without performing the same
analysis here we are going to proceed with the variational formulation for
the present problem following the same line.

Introducing the following bilinear form:%
\begin{eqnarray}
b_{D}(u,v) &:&=-\int_{\Omega}\mathbf{K}\nabla u \cdot \nabla v dxdy
\end{eqnarray}%
\begin{equation}
\begin{array}{l}
b_{J}(u,v):= \\[3mm]
\displaystyle -\int_{\Omega}u \left( \int_{ \mathbb{R} }\left[ u(x+z)-u(x)%
\right] W(dz) \right) dxdy%
\end{array}%
\end{equation}

\bigskip it is possible to define a suitable discretization for our
integro-differential problem \ref{Pricingeq2}.

\begin{remark}
The equation \ref{Pricingeq1} has degenerate coefficients both in the $S$
and in the $y$ variables; while the substitution $x=\log S$ removes the
degeneracy in the $S$ variable, this is still present for the $y$ variable.
In particular we want to discuss the boundary $y=0$. If no restrictions on
the model's parameters would be present, a condition on this boundary should
be imposed in order to have a well-posed problem and the correct condition
is given by \ref{Boundarycond0} . On the other hand, if the restriction on
the parameters of the model given in section 2 \ref{Parameters' restrict}
holds, the variable $y$ never hits that boundary. A closer inspection to the
bilinear form associated to the problem suggests that when Green's lemma is
applied to the l.h.s. of equation \ref{Pricingeq2} an "inflow" condition for
our backward parabolic equation appears if the restrictions on the
parameters hold and this in turn implies that the condition on $y=0$ need
not to be imposed in order to have a well-posed problem.
\end{remark}

\section{Numerical Approach}

In this section we will introduce a Finite Element Discretization of the
above described PIDE. A similar approach for the Merton's and Kou's model
has been introduced in \cite{Almendral}.

\subsection{Temporal Discretization}

The integro-differential equation can be written in the following form
\begin{equation}
\begin{array}{l}
\displaystyle\frac{DF}{Dt}+\nabla\cdot(\mathbf{K}\nabla F)+ \\[3mm]
\displaystyle\int_{-\infty}^{\infty}[F(x+u,y,t)-F(x,y,t)]W(du)-rF=0%
\end{array}%
\end{equation}
where the symbol $\displaystyle\frac{D}{Dt}=\frac{\partial }{\partial t}+(%
\mathbf{a}\cdot\nabla)$ denotes the total derivative.

\vspace{2mm}

The time interval $[0,T]$ will be discretized using a time step $\Delta t$
such that $t^{n+1}=t^n+\Delta t$; moreover we will use the following
notation $F^{n}=F(t^n)$. Starting from this formulation we can obtain the
following temporal discretization
\begin{equation}
\begin{array}{l}
\displaystyle\frac{F^{n+1}-F^n(\tilde{\mathbf{X}})}{\Delta t}+\nabla\cdot(%
\mathbf{K}\nabla F^{n+1})-rF^{n+1}+ \\[3mm]
\displaystyle\int_{-\infty}^{\infty}[F^{n+1}(x+u,y,t)-F^{n+1}(x,y,t)]W(du)=0%
\end{array}%
\end{equation}
where $F^n(\tilde{\mathbf{X}})$ is the value of the price evaluated at the
foot of the characteristic line at time $t^n$ and $\tilde{\mathbf{X}}$ is
the solution of the following initial value problem
\begin{eqnarray}
& &\displaystyle \frac{d \tilde{\mathbf{X}}(\tau;t,\mathbf{x})}{d\tau}=%
\mathbf{a}(\tau;\tilde{\mathbf{X}}(\tau;t,\mathbf{x})) \qquad \tau\in (0,t)
\\
& &\tilde{\mathbf{X}}(t;t,\mathbf{x})=\mathbf{x}.
\end{eqnarray}
This last ODE can be solved using either the implicit Euler method or a more
accurate Runge-Kutta method.

The characteristic Galerkin method, described above, is an
Eulerian-Lagrangian approach and it is stable with a mild stability
criterion allowing therefore the use of a large time step when appropriate.

\subsection{Localization}

The infinite domain has to be reduced to a finite one: in particular we will
consider a rectangular domain $\Omega=[0,x_{max}]\times[0,y_{max}]$. Using
this reduced domain the boundary conditions have to be modifed accordingly
in the following way:
\begin{align}
F(0,y,t)&=0, \\
F(x,0,t)&=\sum_{n=0}^{\infty}e^{-\lambda t}\frac{(\lambda t)^n}{n!}%
F_{BS}(e^x,t,K,\hat{\sigma}_n,\hat{\sigma}_r), \\
F(x_{max},y,t)&=e^{x_{max}}, \\
F(x,y_{max},t)&=e^x.
\end{align}
Moreover the extrema of the integral term have to be reduced to finite
values; to this aim we can use
\begin{equation}
L_{down}=-\sqrt{-2\delta^2 \log ( \epsilon \delta \sqrt{2\pi})} - | \gamma |,
\end{equation}
and
\begin{equation}
L_{up}=\sqrt{-2\delta^2 \log ( \epsilon \delta \sqrt{2\pi})} + | \gamma |.
\end{equation}


\subsection{Spatial Discretization}

The domain $\Omega$ will be discretized using an unstructured triangular
mesh.

The discrete weak formulation reads as follows: find $F_h^{n+1} \in V_h$
such that
\begin{equation}
\begin{array}{l}
\displaystyle \int_{\Omega}F_h^{n+1}\psi \mathrm{d}\mathbf{x} -
\int_{\Omega}F_h^n(\tilde{\mathbf{X}})\psi \mathrm{d}\mathbf{x}+ \Delta t
b_D(F^{n+1}_h,\psi) \\[3mm]
\displaystyle \Delta t b_J( F^{n+1}_h,\psi) - \Delta
t\int_{\Omega}rF_h^{n+1}\psi\mathrm{d}\mathbf{x}=0 \quad \forall \psi \in V_h%
\end{array}%
\end{equation}
where $V_h$ is suitable functional space. In the present paper the unknown $F
$ will be approximated using $\hbox{\Bbt P}_1$ (linear) finite element
\textit{i.e.}
\begin{equation}
F_h^n(x,y)=\sum_{i=1}^{NN}F_i^n\psi_i(x,y),
\end{equation}
where $NN$ is the number of nodes of the triangulation and $\psi_i(x,y)\in%
\hbox{\Bbt P}_1(\Omega)$.

The problem has been solved using FreeFEM++.

\section{Numerical Results}

In this section we shall provide some comments on the numerical results
obtained in order to assess the effectiveness of the proposed numerical
method. In particular we'll present the implicit volatility surfaces for the
following sets of parameters taken from \cite{Thomsen}, where a suitable
calibration methodology has been developed for a large class of stochastic
volatility models with jumps. Moreover we'll provide a graphical comparison
between the solution obtained with our finite element approach and that
obtained by the usual method proposed by P. Carr and D. Madan based on the
Fast Fourier Transform \cite{Carr}.

The range of the strike prices for the european call option considered is $%
80 \leq K \leq 120$ for Fig.1, 3, 4, while for Fig. 2 is $80 \leq K \leq 100$%
. The maturities are between 0 and 3 years. The initial value of the
underlying $S$ has been set $S=100$.

\vspace{0.5cm}

\centerline{
\begin{tabular}{|c||cccc|}
\hline\hline
Set & $\xi$ & $\eta$ & $\theta$ & $\rho$ \\
\hline \hline
$S_1$  & 0.21568 & 0.04937 & 0.23828 & -0.44793  \\
$S_2$  & 0.33502 & 0.033582 & 0.26969 & -0.42404  \\
$S_3$  & 0.13279 & 0.18193 & 0.37518 & -0.59722  \\
$S_4$  & 0.48443 & 0.022097 & 0.21903 & -0.40066 \\
\hline\hline
Set & $k$ & $\delta$ & $\lambda$ & \\
\hline\hline
$S_1$ & -0.11889 & 0.17189 & 0.13674 & \\
$S_2$ & -0.077973 & 0.11048 & 0.33785 & \\
$S_3$ &  0.080396 & 0.057373 & 0.05218 & \\
$S_4$ & -0.12938 & 0.16878 & 0.15977& \\
\hline\hline
\end{tabular}
}

\vspace{0.5cm}

\begin{figure}[tbp]
\center{
\label{S1_1a}
\includegraphics[width=10cm]{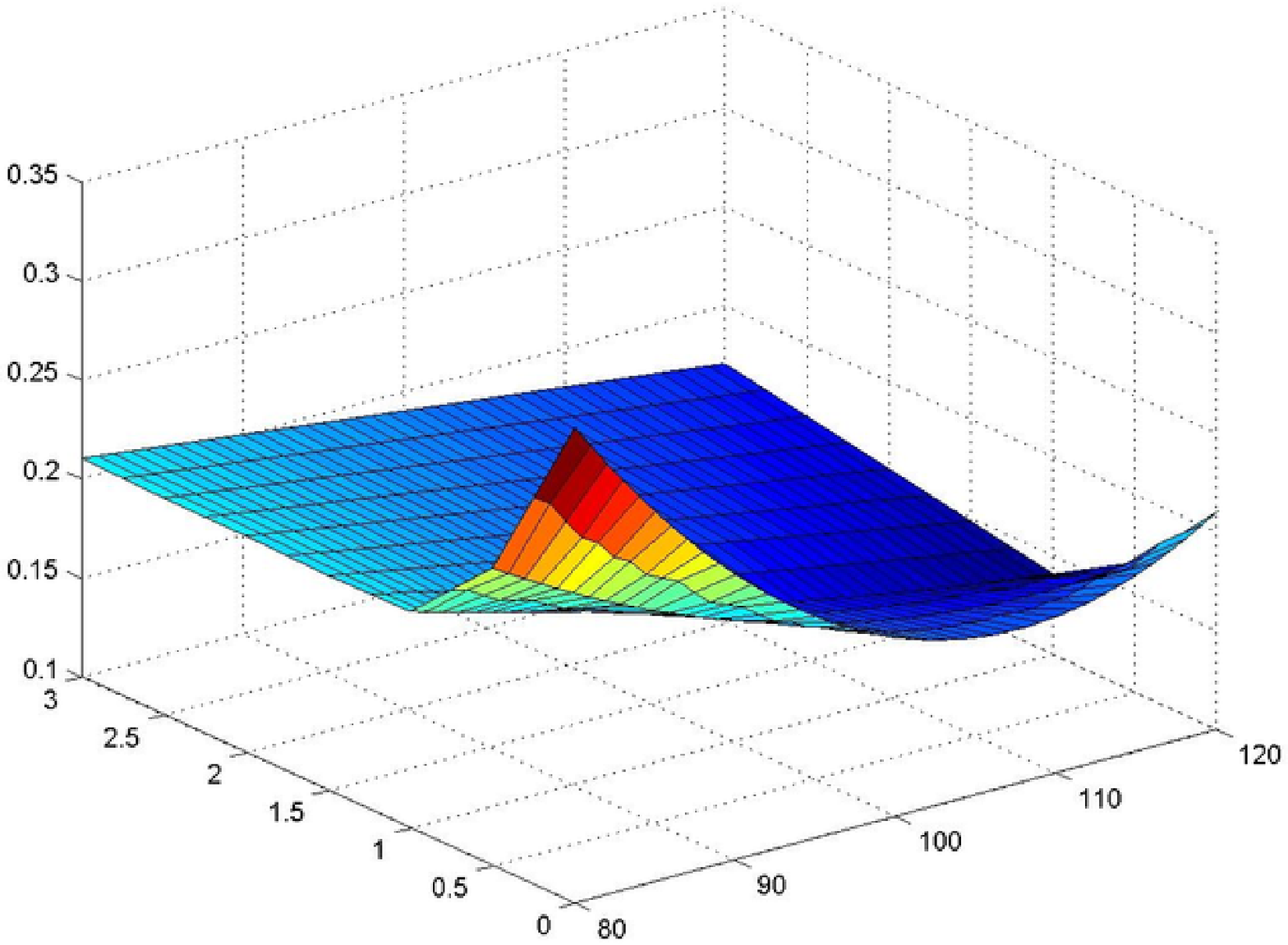}
\caption{Implicit Volatility Surface for parameter set S1.}
}
\end{figure}

\begin{figure}[tbp]
\center{
\label{S2_1a}
\includegraphics[width=10cm]{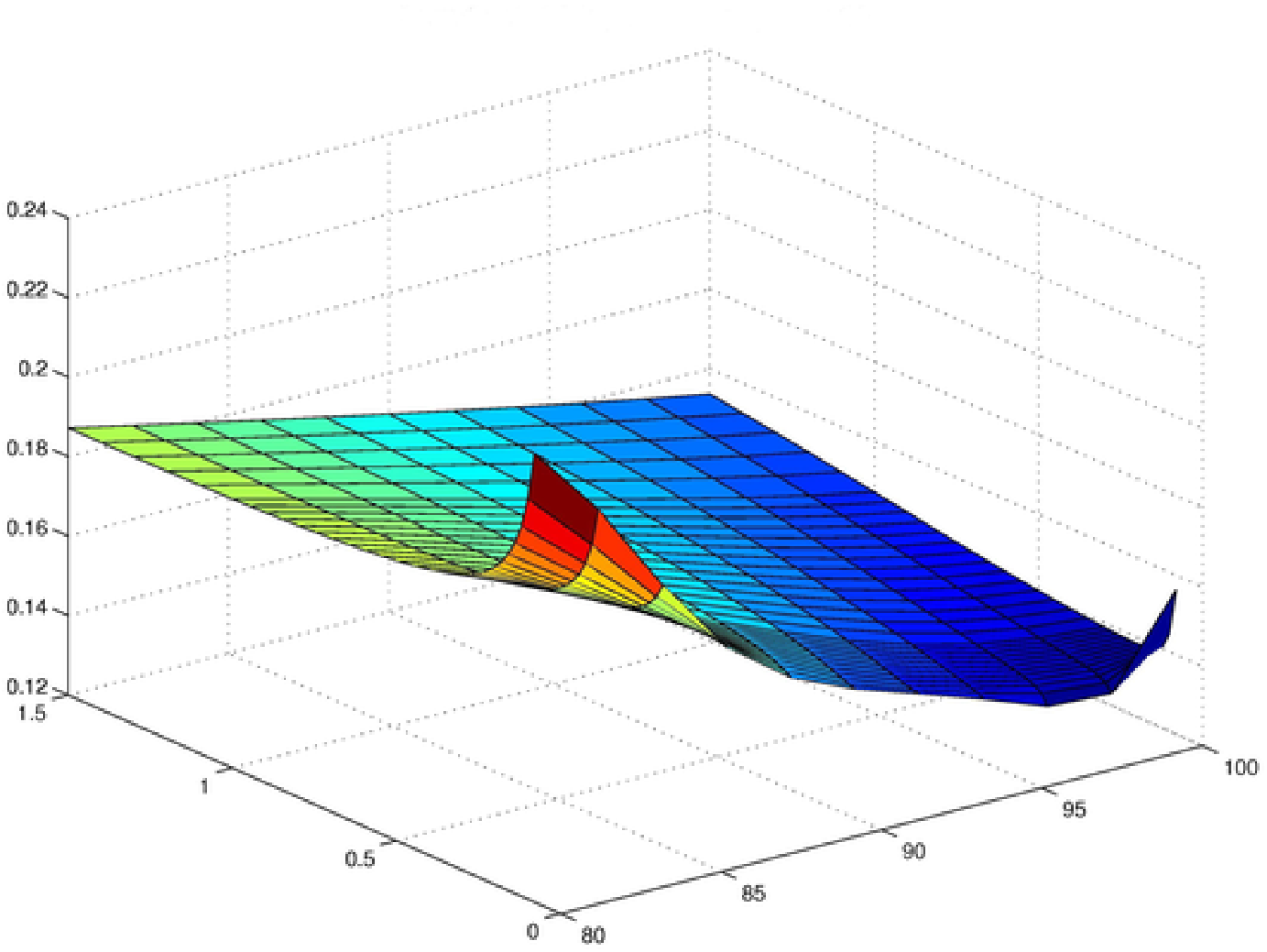}
\caption{Implicit Volatility Surface for parameter set S2.}
}
\end{figure}

\begin{figure}[tbp]
\center{
\label{S3_1a}
\includegraphics[width=10cm]{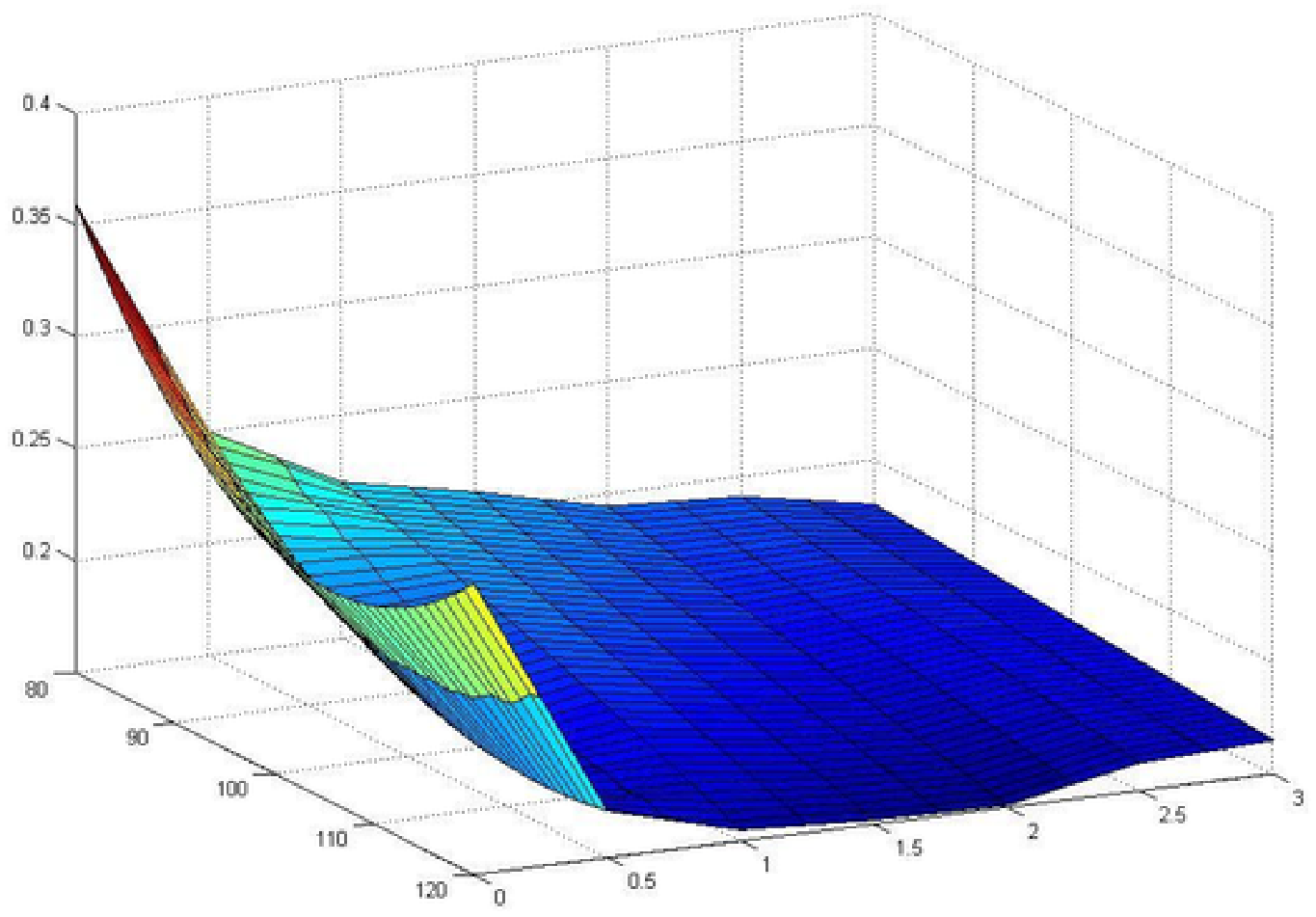}
\caption{Implicit Volatility Surface for parameter set S3.}
}
\end{figure}

\begin{figure}[tbp]
\center{
\label{S4_1a}
\includegraphics[width=8cm]{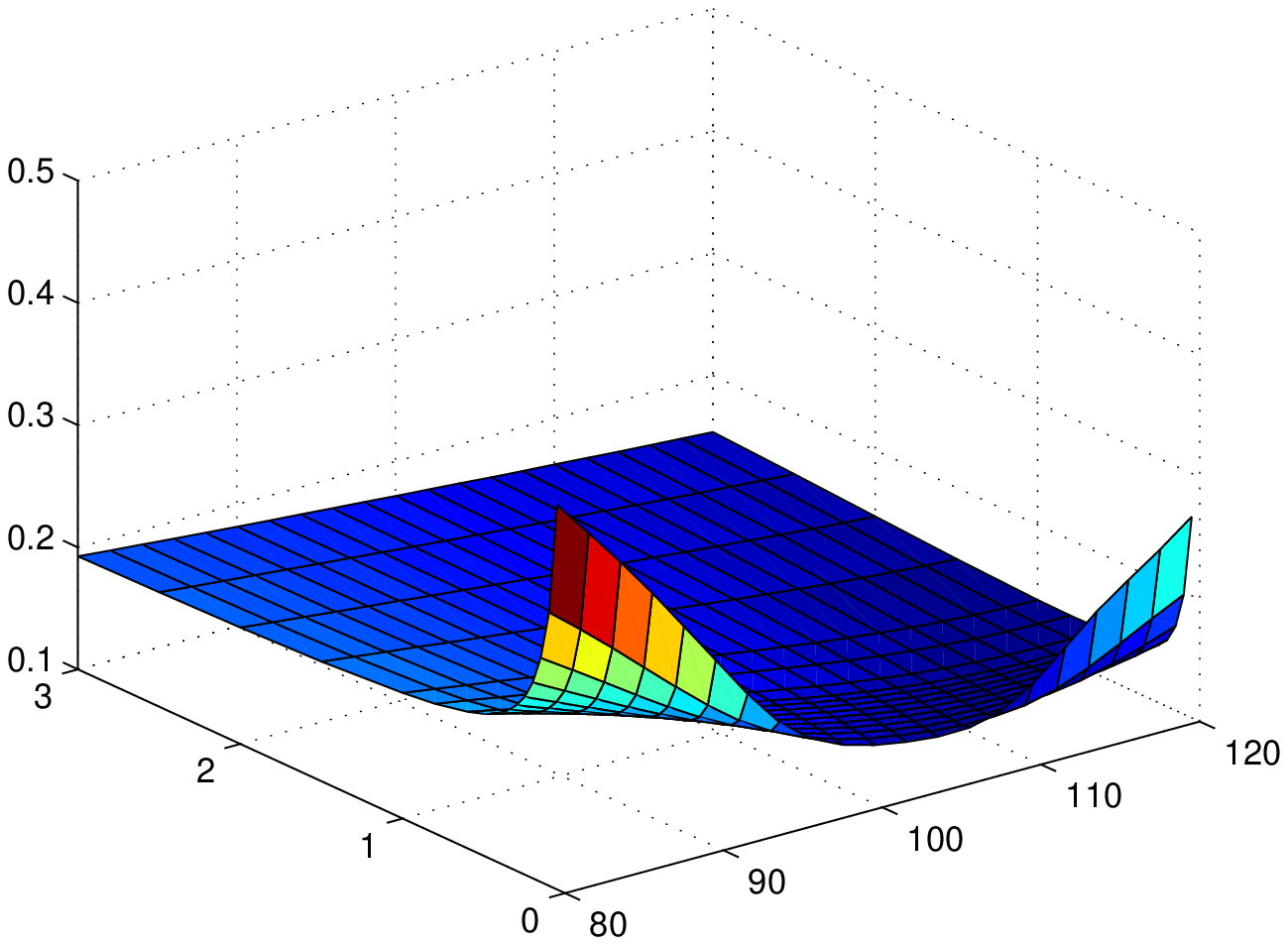}
\caption{Implicit Volatility Surface for parameter set S4.}
}
\end{figure}

All the volatility surfaces obtained exhibit both smiles and skews for short
maturities. While the skew is quite pronounced for the set of parameters
denoted by $S_2$, $S_3$ it seems less evident for the other sets $S_1$, $S_4$%
. The smiles appear less and less pronounced for longer maturities for all
sets. The explanation of the different behavior exhibited by the volatility
surfaces seems to be related more to the different values of the diffusion
coefficient $\theta$ of the volatility dynamics than to the other
parameters, while the "leverage" coefficient $\rho$ seems to be responsible
of both the skews and the smiles appearing in the volatility surfaces. The
behavior of the smiles in correspondence to the "at the money values" of the
call option looks moreover quite realistic.

The next two figures represent the solution obtained with the present finite
element method versus the solution obtained via a Fast Fourier transform
method. While the latter is represented by the continuous line, the former
is indicated by the dots. The set of parameters characterizing the model are
those denoted before by $S_1$, $S_4$ respectively. The range of prices is $%
80 \leq S \leq 120$. The calculation has been performed for $T=1$, $K=100$.

A very good agreement between the solutions can be immediately recognized,
although it looks quite evident that the solution obtained via the finite
element method slightly overprices the call option for higher values of the
underlying at time 0.

In order to check the robustness of the present method with respect to
parameters changed, several trials have been performed corresponding to
other sets of parameters belonging to an enlarged range of parameters and
the results obtained look very close to those just presented.

The CPU time required for a calculation of the call option price for a
single value of the underlying is about 3 minutes, while that required for
the volatility surfaces shown here is about 3 hours on a 2Ghz Centrino Duo
with 1MB RAM. The 2D mesh used in the computation is composed by 2634
triangles and 1398 nodes.

\begin{figure}[tbp]
\center{
\label{FFTS1}
\includegraphics[width=10cm]{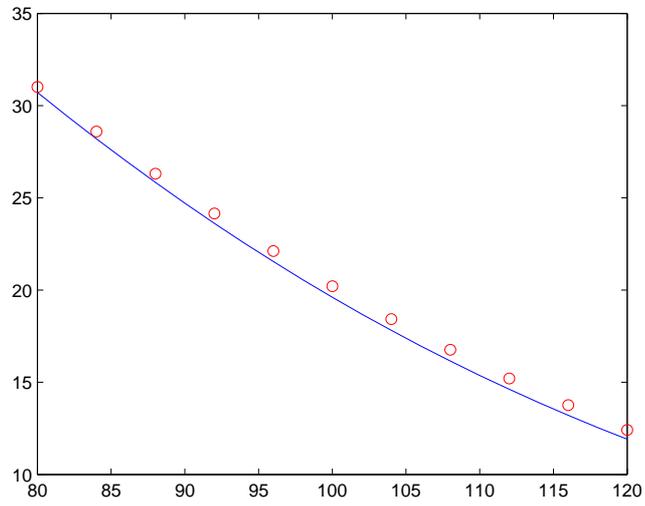}
\caption{Comparison between the FFT solution and the FEM solution for set $S_1$:
$T=1$ and $K=100$.}
}
\end{figure}

\begin{figure}[tbp]
\center{
\label{FFTS4}
\includegraphics[width=10cm]{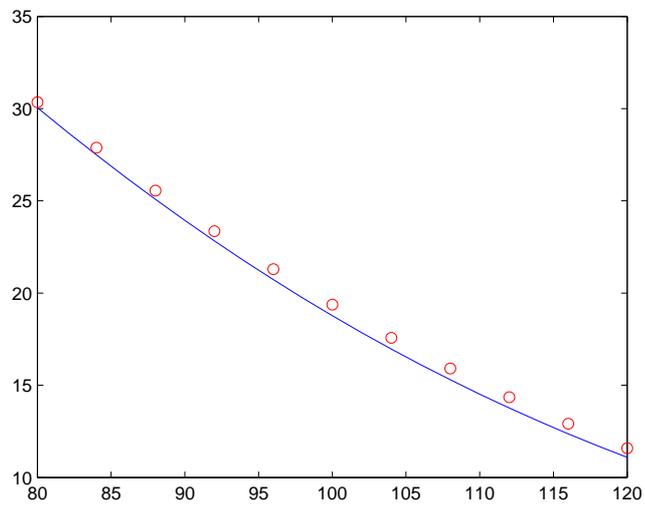}
\caption{Comparison between the FFT solution and the FEM solution for set $S_4$:
$T=1$ and $K=100$.}
}
\end{figure}

\section{Concluding Remarks}

We have presented a finite element approach to the european option valuation
problem formulated via a partial integro-differential equation. Several
choices of the functional space suitable for the spatial discretization are
possible and we have made the most simple choice in order to obtain a fast
and efficient algorithm. Other choices have been made by some authors in
similar contexts, like in \cite{Hilber}, \cite{Matache2}, where Wavelets
have been used in an extensive way to produce an accurate algorithm, but the
convergence of the method with this choice seems to be not very fast. In the
present context, with a single underlying, the choice of piece-wise
polynomial function spaces seems to perform slightly better.

A natural development of the present analysis will be the variational
formulation and the implementation of a finite element method for American
option pricing with the Bates model. This problem can be formulated as a
free boundary problem for the same Partial Integro-Differential equation
given before. The finite element approach seems to be the most promising way
to obtain fast and accurate algorithms for this problem. This will be the
subject of future investigations.

\end{document}